\documentclass[12pt]{amsart}
\usepackage{amssymb,times}
\usepackage{amsmath}
\numberwithin{equation}{section}
\def\ca{{\mathcal A}}

\def\ce{{\mathcal E}}

\def\car{{\mathcal R}}
\def\cs{{\mathcal S}}

\def\cz{{\mathcal Z}}

\def\ga{{\mathfrak A}}
\def\gb{{\mathfrak B}}
\def\gc{{\mathfrak C}}
\def\gd{{\mathfrak D}}

\def\gz{{\mathfrak Z}}

\def\bc{{\mathbb C}}

\def\bm{{\mathbb M}}
\def\bn{{\mathbb N}}

\def\br{{\mathbb R}}

\def\bz{{\mathbb Z}}

\def\a{\alpha}
\def\b{\beta}
\def\g{\gamma}  \def\G{\Gamma}
\def\d{\delta}  
\def\ve{\varepsilon}
\def\e{\epsilon}
 \def\L{\Lambda}

\def\m{\mu}

\def\n{\nu}
\def\r{\rho}
\def\s{\sigma} 
\def\t{\tau}
\def\f{\varphi} \def\F{\Phi}
  \def\Th{\Theta}
\def\om{\omega} \def\Om{\Omega}

\newtheorem{thm}{Theorem}[section]
\newtheorem{lem}[thm]{Lemma}

\newtheorem{prop}[thm]{Proposition}
\newtheorem{defin}[thm]{Definition}

\def\id{\mathop{\rm id}}
\def\tr{\mathop{\rm Tr}}

\def\di{\mathop{\rm d}\!}

\newcommand{\be}{\begin{equation}}
\newcommand{\ee}{\end{equation}}
\newcommand{\bes}{\begin{equation*}}
\newcommand{\ees}{\end{equation*}}
\newcommand{\lb}{\label}
\newcommand{\ds}{\displaystyle}

\newcommand{\nn}{\nonumber}

\begin{document}
\title[Markov states on CAR]
{Markov states on the CAR algebra}
%\thanks{Work partially supported by CNR and INDAM}
\author{Luigi Accardi}
\address{Luigi Accardi\\
Centro Interdisciplinare Vito Volterra\\
II Universit\`{a} di Roma ``Tor Vergata''\\
Via Columbia 2, 00133 Roma, Italy}
\email{{\tt accardi@@volterra.uniroma2.it}}
\author{Francesco Fidaleo}
\address{Francesco Fidaleo\\
Dipartimento di Matematica\\
II Universit\`{a} di Roma ``Tor Vergata''\\
Via della Ricerca Scientifica, 00133 Roma, Italy}
\email{{\tt fidaleo@@mat.uniroma2.it}}
\author{Farruh Mukhamedov}
\address{Farruh Mukhamedov\\
Department of Mechanics and Mathematics\\
National University of Uzbekistan\\
Vuzgorodok, 700095, Tashkent, Uzbekistan} \email{{\tt
far75m@@yandex.ru}}

\begin{abstract}
The program relative to the investigation of quantum Markov states for 
spin chains based on Canonical Anticommutation Relations algebra is carried on. 
This analysis provides a further step for a 
satisfactory theory of quantum Markov processes.
\end{abstract}

\maketitle

\section{introduction}

Recently, the investigation of the Markov property in quantum setting had a 
considerable growth, 
due to various applications to various fields of Mathematics and 
quantum Physics. The reader is referred to \cite{AF1}--\cite{ALV}, 
\cite{F, FM} and the references cited therein, for recent development of 
the theory of quantum stochastic processes and their applications. 

In some of the above mentioned papers, the Markov property for states 
of spin algebras on the standard lattice $\bz^{d}$ is connected with 
properties of the local Radon--Nikodym density matrices describing the 
state under consideration and with the Kubo--Martin--Schwinger (KMS 
for short) 
boundary condition. Hence, properties of Markov states are 
related to properties of local Hamiltonians 
$\{H_{\L}\}_{\L\subset\bz^{d}}$ canonically associated to that state. 

It should be noted that the usual spin algebra on $\bz^{d}$ satisfies 
the commutation rule for observables localized in separated regions. 

On the other hand, properties of states on quasi--local algebras satisfying 
the Canonical Anticommutation Relations (CAR for 
short) for separated regions are 
investigated in a sequel of recent papers \cite{AM1}--\cite{AM4}, 
that is when Fermion operators are present. The states considered in 
the last papers are (the generalization of) product states with 
respect to a fixed partition of the lattice $\bz^{d}$. Also in this 
case, connections with local interaction, Hamiltonians and with the 
KMS property are established.

In the present paper, the program relative to the investigation of quantum Markov 
states for 
spin chains based on Canonical Anticommutation Relations algebra is carried on. 

As a first step, we restrict the matter to the simplest case when 
there is only one degree of freedom in each site $k$ of a completely 
ordered lattice (i.e. when we have a chain which is a subset of $\bz$, 
and when the CAR algebra in the site $k$ is the full matrix 
algebra $\bm_{2}(\bc)$. We then consider non homogeneous states 
satisfying an appropriate version of the Markov property. In this 
situation, the structure of such Markov states on CAR algebra is 
fully understood. Furthermore, the connection of Markov states on
on CAR algebra and local Hamiltonians is established.

This analysis provides a first step in order to understand the 
structure of Markov states on more complicated CAR algebras on the 
chain, and also on the multidimensional lattice $\bz^{d}$. It 
provides also further step towards a 
satisfactory theory of quantum Markov processes. 

\medskip

Here, for the sake of completeness, we report some preliminary facts which are 
useful in the sequel.

By a (Umegaki) {\it conditional expectation} 
$E:\ga\mapsto\gb\subset\ga$ we mean a norm--one projection of the 
$C^{*}$--algebra $\ga$ onto a $C^{*}$--subalgebra (with the same 
identity $I$) $\gb$.  The map $E$ is automatically a completely 
positive identity--preserving $\gb$--bimodule map, see \cite{S}, 
Section 9.  When $\ga$ is a matrix algebra, the structure of a 
conditional expectation is well--known, see \cite{H} (see also \cite{FI},
for more general cases when the centre 
of the range of $E$ is infinite--dimensional and atomic). Let $\ga$ 
be a full matrix algebra and consider the (finite) set 
$\{P_{i}\}$ of minimal central projections of the range $\gb$ of $E$, 
we have
\be
\lb{condd}
E(x)=\sum_{i}E(P_{i}xP_{i})P_{i}\, .
\ee

Then $E$ is uniquely determined by its values on
the reduced algebras
$$
\ga_{P_{i}}:=P_{i}\ga P_{i}=N_{i}\otimes\bar N_{i}
$$
where $N_{i}\sim\gb_{P_{i}}:=\gb P_{i}$ and 
$\bar N_{i}\sim\gb'_{P_{i}}:=\gb'P_{i}$.\footnote{The commutant $\gb'$ 
is considered in the ambient algebra $\ga$.}  
In fact, there exist states $\phi_{i}$ on $\bar N_{i}$ such that
\begin{equation}
\label{cond}
E(P_{i}(a\otimes\bar a)P_{i})=\phi_{i}(\bar a)P_{i}(a\otimes I)P_{i}\,.
\end{equation}

Consider a triplet $\gc\subset\gb\subset\ga$ of  
unital $C^{*}$--algebras. A {\it quasi--conditional expectation} w.r.t. the given
triplet, is a  completely positive, identity
preserving map $E:\ga\mapsto\gb$ such that
\be
\lb{qce}
E(ca)=cE(a)\,, \quad a\in\ga\,,\, c\in\gc\,.\footnote{Notice that, as 
the quasi--conditional expectation $E$ is a real map, we have
$$E(ac)=E(a)c\,, \quad a\in\ga\,,\, c\in\gc$$ as well.}
\ee

A pivotal example of quasi--conditional expectation is given by the 
$\f$--conditional expectation $E^{\f}:\ga\mapsto\gb$ preserving the 
restriction to the $W^{*}$--subalgebra $\gb$ of a
normal faithful state $\f$ on the $W^{*}$--algebra $\ga$. One can 
choose for $\gc$ any unital $C^{*}$--subalgebra of the $W^{*}$--algebra $\gb$
contained in the $E^{\f}$--fixed point algebra, see \cite{AC}.

\section{Markov States}

Let $\ga$ be the {\it Canonical Anticommutation Relations} (CAR, for short)
algebra, with generating element $a_i$ and their adjoints 
$a^*_i$, $i\in I$. In our situation, the index set $I$ is a
totally ordered countable discrete set $I$
containing, possibly a smallest element $j_{-}$ and/or a greatest
element $j_{+}$. Namely, if $I$ contains neither $j_{-}$, nor
$j_{+}$, then $I\sim\bz$. If just $j_{+}\in I$, then $I\sim\bz_{-}$,
whereas if only $j_{-}\in I$, then $I\sim\bz_{+}$. Finally, if both
$j_{-}$ and $j_{+}$ belong to $I$, then $I$ is a finite set and the 
analysis becomes easier. If  
$I$ is order--isomorphic to $\bz$, $\bz_{-}$ or $\bz_{+}$, we put
simbolically $j_{-}$ and/or $j_{+}$ equal to $-\infty$ and/or $+\infty$
respectively. In such a way, the objects with indices
$j_{-}$ and $j_{+}$ will be missing in the computations.

The generators $\{a_{j}, a^{+}_{j}\}_{j\in I}$ satisfy the relations  
\bes
\{a^{+}_{j},a_{k}\}=\d_{jk}\,,\,\,
\{a_{j},a_{k}\}=\{a^{+}_{j},a^{+}_{k}\}=0\,,\,\,j,k\in I\,,
\ees
where $\{\,\cdot\,,\,\cdot\,\}$ stands for the anticommutator.
The parity automorphism of $\ga$ is denoted by
$\Th$. For any subset $\L\subset I$, the
$C^*$-subalgebra of $\ga$ generated by $a_j, a^{+}_j$ for $j\in\L$ is 
denoted by $\ga_{\L}$. It is well--known that $\ga$ is a 
$\bz_{2}$--graded algebra, $\Th$ being the grading automorpism.
It is well known that the CAR algebra is isomorphic to the $C^{*}$--(infinite) 
tensor product ${\ds\overline{\bigotimes_{I}\bm_{2}(\bc)}^{C^*}}$.
For the basic properties of CAR, we refer the 
reader to \cite{AM2, BR1, T3} and the references cited therein. 

In order to avoid technicalities, we deal only with locally faithful states.
Furthermore, in order to treat translation invariant or periodic 
states,\footnote{Relatively to this point, see \cite{AM2}, and 
\cite{BR2}, Example 5.2.21.}
we consider only $\Th$--invariant (quasi--)conditional expectations, if it is not 
otherwise specified.
\begin{defin}
\lb{dema}
A state $\f$ on $\ga$ is called a Markov state
if, for each $j_{-}\leq j<j_{+}$, there exists a quasi--conditional expectation
$E_{n}$ w.r.t. the triplet 
$\ga_{n-1]}\subset\ga_{n]}\subset\ga_{n+1]}$ satisfying
\begin{align*} 
&\f_{n+1]}\circ E_{n}=\f_{n]}\,,\\
&E_{n}(\ga_{[n,n+1]})\subset\ga_{\{n\}}\,.
\end{align*}
\end{defin}

We show that the Markov property defined above can be stated by 
a sequence of global quasi--conditional expectations, or equally well 
by sequences ol local or global conditional expectations.
\begin{prop}
\lb{twst}
Let $\f$ be a state on the CAR algebra. The following assertions are 
equivalent.
\begin{itemize}
\item[(i)] $\f$ is a Markov state;
\item[(ii)] the properties listed in Definition \ref{dema} are 
satisfied if we replace the quasi--conditional expectations $E_{n}$ 
with Umegaki conditional expectations $\ce_{n}$;   
\item[(iii)] for each $j<j_{+}$, there exists a conditional expectation
$\ce_{n]}:\ga\mapsto\car(\ce_{n]})\subset\ga_{n]}$ satisfying
\begin{equation*} 
\f\circ\ce_{n]}=\f\,,
\end{equation*}
\begin{equation}
\lb{adcd}
\ce_{n]}(\ga_{[n})\subset\ga_{\{n\}}\,;
\end{equation}
\item[(iv)] the properties listed in (iii) are satisfied if we replace the
conditional expectations $\ce_{n]}$ with quasi--conditional expectations
$E_{n]}$
\end{itemize}
\end{prop}
\begin{proof}
It is enough to prove (i)$\Rightarrow$(ii) and (ii)$\Rightarrow$(iii), 
the remaining implications being trivial.

\noindent
(i)$\Rightarrow$(ii) Consider the restriction 
$e_{n}:=E_{n}\lceil_{\ga_{[n,n+1]}}$ which is a completely 
positive, identity preserving map leaving invariant a faithful state.
Taking the ergodic limit
$$
\ve_{n}:=\lim_k\frac{1}{k}\sum^{k-1}_{h=0}(e_{n})^{h}\,,
$$
we provide a conditional expectation leaving invariant (the 
restriction of) the state $\f$. In order to define a conditional 
expectation on $\ga_{n+1]}$, we start by noticing that 
$a\in\ga_{n+1]}$ can be written in a unique way as
\be
\lb{modu}
a=\sum 
c_{(j_{n},j_{n+1})(k_{n},k_{n+1})}e(n)_{j_{n}k_{n}}e(n+1)_{j_{n+1}k_{n+1}}\,,
\ee
where $c_{(j_{n},j_{n+1}),(k_{n},k_{n+1})}\in\ga_{n-1]}$, and 
the $e(n)_{j_{n}k_{n}}e(n+1)_{j_{n+1}k_{n+1}}$ given in \cite{T3}, 
pag. 92, 
provide a system of matrix units for $\ga_{[n,n+1]}$.
We define for $a\in\ga_{n+1]}$ written as in \eqref{modu},
$$
\ce_{n}(a):=\sum 
c_{(j_{n},j_{n+1}),(k_{n},k_{n+1})}\ve_{n}(e(n)_{j_{n}k_{n}}e(n)_{j_{n+1}k_{n+1}})\,.
$$

As a generic $a_{i}\in\ga_{n+1]}$ has the form $a_{i}=\sum c_{i,\a}e_{\a}$,
we compute
$$
\ce_{n}(a^{*}_{i}a_{j})=\sum 
c^{*}_{i,\a}\ve_{n}(e^{*}_{\a}e_{\b})c_{j,\a}\,,
$$
where the equality follows taking into account that
$\ve_{n}$ is $\Th$--invariant. This means that $\ce_{n}$ is completely 
positive (see \cite{T1}, Section IV.3). Namely, we get a norm one 
projection of $\ga_{n+1]}$ onto a $*$--subalgebra of $\ga_{n]}$ 
satisfying all property listed in Definition \ref{dema}.

\noindent
(ii)$\Rightarrow$(iii) Let $m>n$, define
$$
\ce_{n,m}:=\ce_{n}\circ\cdots\circ\ce_{m-1}\,.
$$

As $\ce_{n,m+k}\lceil_{\ga_{m-1]}}=\ce_{n,m}\lceil_{\ga_{m-1]}}$, the 
direct limit
$$
\ce_{n}^{0}:=\lim_{\stackrel{\longrightarrow}{m\uparrow j_{+}}}\ce_{n,m}
$$
is a well defined norm one projection of the dense subalgebra
${\ds\bigcup_{m}\ga_{m]}}$ onto a subalgebra of $\ga_{n]}$ which, by 
continuity, uniquely extends to a conditional expectation 
satisfying the required properties.
\end{proof}

An immediate application of Proposition \ref{twst} is that the Markov 
state $\f$ satisfies all the properties listed in Definition 6.1 of \cite{AF3}.
Indeed, it is sufficient to put inside $I$, $\a:=[n+1$, $\bar\a=[n$. 
In such a way, $\a'=n]$, $\bar\a'=n-1]$ and the projective net of 
conditional expectations is precisely that formed by the 
expectations $\ce_{n]}$ given in (iii) of Proposition \ref{twst}.
However, in order to investigate further structural properties of 
Markov states when Fermions are present, the $\Th$--invariance for the 
conditional expectations and 
the additional condition \eqref{adcd} are needed, see below.

As is stated in Proposition \ref{twst}, the main object is the 
$\Th$--invariant two--step conditional expectation $\ve_{n}$. So, we 
should describe all $\Th$--invariant subalgebras of $\ga_{\{n\}}=\bm_{2}(\bc)$. 
Of course, 
$\bc I$ and $\bm_{2}(\bc)$ are trivially $\Th$--invariant. It remains 
open the case when the $\Th$--invariant subalgebra is a maximal abelian 
subalgebra of $\bm_{2}(\bc)$.
\begin{lem}
\lb{mc}
The unique $\Th$--invariant maximal abelian subalgebra of the CAR 
algebra generated by $a, a^{+}$ is generated by the projection
$aa^{+}$ and $a^{+}a$.
\end{lem}
\begin{proof}
Let $P$ one of the minimal projection generating the algebra under 
consideration. Then $P=P_{+}+P_{-}$ is its splitting in the even and 
odd part. Then, $\Th(P)$ is another minimal projection in the same 
maximal abelian subalgebra. This means that $\Th(P)=I-P$, which is 
excluded as this implies $I=2P_{+}$. The remaining possibility is $\Th(P)=P$, which is 
equivalent to $P=P_{+}$. The last assertion turns out to be equivalent to 
$P=aa^{+}$, or $P=a^{+}a$ which is the assertion.
\end{proof}
\begin{lem}
\lb{mc1}
If $\car(\ve_{n})=\ga_{\{n\}}$ then
$$
\ve_{n}(\ga_{\{n+1\},-})=0\,.
$$
\end{lem}
\begin{proof} 
If $x_{n+1}\in\ga_{\{n+1\}}$ is odd,
then $x_{n+1}$ anticommutes with $a_{n},a^+_n$.
Hence, $\ve_{n}(x_{n+1})$ anticommutes with $a_{n},a^+_n$ as well.
As
$$
\ve_{n}(x_{n+1})=\a a_n^++\b a_n+\g a_na^+_n+\d a^+_na_n\,,
$$
we have
\begin{align*}
&a_n\ve_{n}(x_{n+1})=\a a_na^+_n+\g a_n\,,\\
&\ve_{n}(x_{n+1})a_n=\a a^+_na_n+\d a_n\,.
\end{align*}

Using the above anticomutation properties, we infer that
$$
\a(a_na^+_n+a_n^+a_n)+(\g+\d)a_n=0,
$$
which implies $\a=0$ and $\d=-\g$.

By the similar argument applied to $a^{+}_{n}$, we get
$\b=0$. Thus,
$$
\ve_{n}(x_{n+1})=\g(a^+_na_n-a_na^+_n)
$$
which means $\g=0$ as $\ve$ is supposed to be $\Th$--invariant.
\end{proof}

We pass to the study of the structure of the $\ve_{n}$ for the three 
possibilities listed below (see Lemma \ref{mc}). This should 
be done with some caution, as \eqref{condd}, \eqref{cond} do not directly apply to 
our situation. According to the standard terminology reported in pag. 92 of 
\cite{T3}, we put
\begin{align*}
&P_{1}^{n}:=a_{n}a_{n}^{+}\equiv e_{11}(n)\,,\\
&P_{2}^{n}:=a_{n}^{+}a_{n}\equiv e_{22}(n)\,.
\end{align*}

\begin{prop}
\lb{stce}
Under the above assumptions, the following assertions hold true.
\begin{itemize}
\item[(i)] If $\car(\ve_{n})=\bc I$, then there exists a even state 
$\F_{n}$ on $\ga_{[n,n+1]}$ such that $\ve_{n}(x)=\F_{n}(x)I$;
\item[(ii)] If $\car(\ve_{n})=\ga_{\{n\},+}$ then there exist even 
states $\F^{n}_{1}$, $\F^{n}_{2}$ on $\ga_{\{n+1\}}$ such that, for 
$x\in\ga_{\{n\}}$, $y\in\ga_{\{n+1\}}$,
\bes
\ve_{n}(xy)=\tr(xP_{1}^{n})\F^{n}_{1}(y)P_{1}^{n}+\tr(xP_{2}^{n})\F^{n}_{2}(y)P_{1}^{n}\,;
\ees
\item[(iii)] $\car(\ve_{n})=\ga_{\{n\}}$ then there exists a even state 
$\Psi_{n}$ on $\ga_{\{n+1\}}$ such that, for 
$x\in\ga_{\{n\}}$, $y\in\ga_{\{n+1\}}$, $\ve_{n}(xy)=x\Psi_{n}(y)$.
\end{itemize}
\end{prop}
\begin{proof}
(i) and (ii) easily follow by \eqref{condd}, \eqref{cond}, taking into 
account that 
$\ga_{\{n\},+}\bigvee\ga_{\{n+1\}}\sim\ga_{\{n\},+}\otimes\ga_{\{n+1\}}$ 
(\cite{T3}, pag. 94), and $\Th$--invariance of $\ve_{n}$.

\noindent
(iii) By a repeated application of Lemma \ref{mc1}, 
if $x\in\ga_{\{n\}}$, $y\in\ga_{\{n+1\}}$, we have
\bes
x\ve_{n}(y)=x\ve_{n}(y_{+})
=\ve_{n}(xy_{+})=\ve_{n}(y_{+}x)
=\ve_{n}(y_{+})x=\ve_{n}(y)x\,.
\ees

This means that $\ve_{n}(y)\in\cz(\ga_{\{n\}})\equiv\bc I$. The 
assertion follows again by $\Th$--invariance of $\ve_{n}$.
\end{proof}

It is immediate to verify that $\F_{n}$, $\Psi_{n}$ are the 
restrictions of $\f$ to $\ga_{[n,n+1]}$, $\ga_{\{n+1\}}$ 
respectively, and
$$
\F^{n}_{i}=\frac{\f(P^{n}_{i}\,\cdot\,)}{\f(P^{n}_{i})}\,,\quad i=1,2\,.
$$

We leave to the reader the proof of the following 
\begin{lem}
\label{propr}
Let $\f$ be a Markov state on the CAR algebra, and $\{\ve_{j}\}_{j_{-}\leq j<j_{+}}$
the associated sequence of two--point conditional expectations. Then
\begin{equation}
\label{3}
\f(x_{k}\cdots x_{l})
=\f((\ve_{k}(x_{k}\ve_{k+1}(x_{k+1}\cdots\ve_{l-1}(x_{l-1}
x_{l})\cdots)))
\end{equation}
for every $k,l\in I$ with $k<l$, and
$x_{k}x_{k+1}\cdots x_{l-1}x_{l}$
any linear generator of $\ga_{[k,l]}$.
\end{lem}

Now we show that a Markov state can be obtained by lifting, via a 
suitable conditional expectation, its restriction to a subalgebra. 
This property is analogous to the corresponding one for Markov states 
on spin chains (see \cite{ALi}), and plays a crucial r\^ole in the 
sequel (see also \cite{AF1}, Section 3).

We start by defining a conditional expectation onto a subalgebra of 
the CAR algebra $\ga$. Let $\G\subset I\backslash\{j_{+}\}$ be defined 
as the set of sites $n$ such that $\car(\ve_{n})=\ga_{\{n\},+}$. Define
${\ds\ce:\ga\mapsto\overline{\ga_{I\backslash\G}\bigvee
\big(\bigvee_{n\in\G}\ga_{\{n\},+}
\big)}^{C^{*}}}$ as follows. Put
\be
\lb{ceinv}
\ce:=\prod_{j\in I}F_{j}\,,
\ee
where $F_{j}$ is the identity if $j\notin\G$, and 
$$
F_{j}(x)=P_{1}^{j}xP_{1}^{j}+P_{2}^{j}xP_{2}^{j}
$$
otherwise. The map $\ce$ is well defined on localized elements
and extends by continuity to a conditional expectation on 
$\ga$ onto ${\ds\overline{\ga_{I\backslash\G}\bigvee
\big(\bigvee_{n\in\G}\ga_{\{n\},+}
\big)}^{C^{*}}}$.
\begin{prop}
Let $\f$ be a Markov state, and $\ce$ the conditional expectation 
defined in \eqref{ceinv}. Then $\f=\f\circ\ce$.
\end{prop}
\begin{proof}
Taking into account \eqref{condd}, we get if $n\in\G$,
\begin{equation*}
\ve_{n}(xy)=\sum_{k=1}^{2}\ve_{n}(P_{k}^{n}xyP_{k}^{n})P_{k}^{n})=
\ve_{n}\bigg(\sum_{k=1}^{2}P_{k}^{n}xP_{k}^{n}y\bigg)=\ve_{n}(\ce(x)y)\,.
\end{equation*}

Hence, by Lemma \ref{propr} we obtain for $k<l<j_{+}$, and $x_{j}$ 
linear generators of $\ga$, 
\begin{align*}
&\f(x_{k}\cdots x_{l})
=\f((\ve_{k}(x_{k}\ve_{k+1}(x_{k+1}\cdots\ve_{l}(x_{l})\cdots)))\\
&=\f((\ve_{k}(\ce(x_{k})\ve_{k+1}(\ce(x_{k+1})\cdots\ve_{l}(\ce(x_{l})\cdots)))
=\f(\ce(x_{k})\cdots\ce(x_{l}))
\end{align*}
which leads to the assertion.
\end{proof}

\section{the structure of Markov states}

In this section we provide a disintegration of a Markov state into elementary
Markov states. This allows us to give a reconstruction theorem. 
These results parallels the analogous one described in \cite{AF1}. 

We start by partitioning $I\backslash\{j_{+}\}$ into disjoint intervals 
each of which consisting of points $n$ such that $\car(\ve_{n})$ is 
trivial (i.e. $\bc I$ or $\ga_{\{n\}}$), or $\car(\ve_{n})=\ga_{\{n\},+}$. 
In this way, ${\ds\G=\stackrel{\circ}{\cup}_{k}\G_{k}}$ (where 
$\stackrel{\circ}{\cup}$ stands for disjoint union), and 
$\G_{k}=(l_{k}-1,r_{k}+1)$. 

Define
\begin{equation}
\label{compact}
\Om:=\prod_{k}\Om_{k}\,,\quad\Om_{k}:=\prod_{l_{k}-1< j<
r_{k}+1}\{1,2\}\,,\quad\m:=\prod_{k}\m_{k}\,,
\end{equation}
where $\m_{k}$ is the Markov measure on $\Om_{k}$ determined by the 
distributions $\pi^{j}_{\om_{j}}$ at place $j$ and the transition 
coefficients 
$\pi^{j}_{\om_{j}\om_{j+1}}$ given by
\begin{align}
\label{trp}
\pi^{j}_{\om_{j}}&=\f(P^{j}_{\om_{j}})\,,\quad l_{k}-1<j<r_{k}+1\,,\,\,
\om_{j}=1,2\,,\\ 
\pi^{j}_{\om_{j}\om_{j+1}}&=\frac{\f(P^{j}_{\om_{j}}P^{j+1}_{\om_{j+1}})}
{\f(P^{j}_{\om_{j}})}\,,
\quad l_{k}-1<j<r_{k}\,,\,\,\om_{j},\om_{j+1}=1,2\nn\,.
\end{align}

Notice that the range of $\ce$ given in \eqref{ceinv} can be 
described by the $C^{*}$--algebra consisting of all continuous 
functions $\om\in\Om\mapsto x(\om)\in\ga_{I\backslash\G}$. Furthermore,
the measure $\m$ is precisely given, under a standard isomorphism, by 
the restriction of the Markov state $\f$ to the Abelian $C^{*}$--subalgebra 
generated by the projections $\{P^{j}_{\om_{j}}\,\big|\, j\in\G\,,\,\om_{j}=1,2\}$.

Starting from the Markov state $\f$, consider, for $\om\in\Om$ the product state 
extension (product state for short, see 
\cite{AM3})
\be
\lb{pse}
\psi_{\om}=\prod_{k}\psi_{k,\om}\,,
\ee
on $\ga_{I\backslash\G}$. Here, $\psi_{k,\om}$ is the one--step or 
two--step product state on 
$\ga_{(r_{k},l_{k+1})}$ depending only on $\om_{r_{k}}$, 
$\om_{l_{k+1}}$,
constructed as follows.\footnote{If $k+1$ is the 
first element of $\G$ not equal to $j_{-}$, then $r_{k}=j_{-}$. If 
$k$ is the last element of $\G$, then $l_{k+1}=j_{+}$. We 
are using also intervals without the boundary elements in order to 
take into account the possibility of $j_{-}=-\infty$ and/or $j_{+}=+\infty$.}
\begin{itemize}
\item[(i)] If $k+1$ is the 
first element of $\G$ not equal to $j_{-}$, or $\car(\ve_{r_{k}+1})=\bc I$, then
$$
\psi_{k,\om}(x):=\frac{\f(xP^{l_{k+1}}_{\om_{l_{k+1}}})}{\f(P^{l_{k+1}}_{\om_{l_{k+1}}})}\,,
$$
\item[(ii)] if $k$ is the last element of $\G$, or 
$\car(\ve_{l_{k+1}-1})=\ga_{\{l_{k+1}-1\}}$, then 
$$
\psi_{k,\om}(x):=\frac{\f(P^{r_{k}}_{\om_{r_{k}}}x)}{\f(P^{r_{k}}_{\om_{r_{k}}})}\,,
$$
\item[(iii)] if the interval under consideration has on the left and 
on the right, elements of $\G$, that is it has the form
$[r_{k}+1,l_{k+1}-1]$, then 
$$
\psi_{k,\om}(x):=\frac{\f(P^{r_{k}}_{\om_{r_{k}}}xP^{l_{k+1}}_{\om_{l_{k+1}}})}
{\f(P^{r_{k}}_{\om_{r_{k}}})\f(P^{l_{k+1}}_{\om_{l_{k+1}}})}\,,
$$
\item[(iv)] for $r_{k}<j<l_{k+1}-1$, the two--step factor 
$\psi_{k,\om}(x)$, $x\in\ga_{[j,j+1]}$ 
appears iff $\car(\ve_{j})=\bc I$ and 
$\car(\ve_{j+1})=\ga_{\{j+1\}}$.
\end{itemize}

Notice that, by Proposition 
\ref{stce}, the states $\psi_{\om}$ are even.
Finally, it is easy to show that the map
\be
\lb{mesur}
\om\in\Om\mapsto\psi_{\om}\in\cs(\car(\ce))
\ee
is measurable in the weak--$*$ topology.

We are ready to prove a result concerning the decomposition of
a Markov state on the CAR algebra into elementary Markov states 
following the strategy developed in Section 3 of \cite{AF1}.
\begin{thm}
\label{dis}
Let $\f$ be a Markov state on the CAR algebra $\ga$.

Then $\f$
admits a direct--integral decomposition
\be
\lb{didec}
\f=\int^{\oplus}_{\Om}\psi_{\om}(\ce(\,\cdot\,)(\om))\m(\di\om)\,,
\ee
where the measure space $(\Om,\m)$ is defined in \eqref{compact}, \eqref{trp}, 
the conditional expectation $\ce$ is given in \eqref{ceinv}, the 
state $\psi_{\om}$ is given in \eqref{pse} through (i)--(iv) above, and 
finally the integral \eqref{didec} is understood as a $L^{1}$--direct 
integral.\footnote{See \cite{T1}, Section IV.8.}
\end{thm}
\begin{proof}
We sketch the proof which is quite similar to that of Theorem 3.2 of 
\cite{AF1}. 

Put $\gb:=\car(\ce)$. 
Consider the abelian $C^{*}$--subalgebra $\gz$ of $\gb$ generated by 
$P^{j}_{i}$, $j_{-}\leq j<j_{+}$, $i=1,2$, 
together with
the GNS representation $\pi$ of $\gb$ relative to $\f_{\lceil\gb}$. 
Then 
$\pi(\gz)''\subset\pi(\gb)'\cap\pi(\gb)''$. As
$\pi(\gz)''\sim L^{\infty}(\Om,\m)$, we have for
$\pi$ the direct--integral disintegration
$$
\pi=\int^{\oplus}_{\Om}\pi_{\om}\m(\di\om)
$$
where $\om\mapsto\pi_{\om}$ is a weakly measurable field of representations of
$\gb$, see \cite{T1}, Theorem IV.8.25.

Further, by mimicking the proof of Proposition IV.8.34 of \cite{T1},
we find a measurable field $\om\mapsto\xi_{\om}$ of vectors such
that, for each $x\in\gb$, we get
$$
\f(x)=\int^{\oplus}_{\Om}\langle\pi_{\om}(x)\xi_{\om},\xi_{\om}\rangle\m(\di\om)\,.
$$

As $\f$ is a Markov state, it is invariant w.r.t. $\ce$. Then
$$
\f=\int^{\oplus}_{\Om}\f_{\om}\m(\di\om)
$$
for the measurable field $\f_{\om}$ defined as
$$
\f_{\om}:=\langle\pi_{\om}(\ce(\,\cdot\,))\xi_{\om},\xi_{\om}\rangle\,.
$$

Fix localized elements $x\in\ga$, $z\in\gz\sim C(\Om)$. It is easy to 
show by applying the Markov property, that
$$
\int_{\Om}z(\om)\f_{\om}(x)\m(\di\om)=
\int_{\Om}z(\om)\psi_{\om}(E_{\om}(x))\m(\di\om)
$$
for each fixed localized operator $x\in\ga$ and each 
$z\in\gz$ represented by the function $z(\om)$ in $C(\Om)$ depending only on 
finitely many variables. As such
functions are
dense in $C(\Om)$, we conclude
by the uniqueness of the Radon--Nikodym derivative, that for each localized
element $x\in\ga$, there exists
a measurable set $\Om_{x}\subset \Om_{0}$ of full $\m$--measure such that,
when $\om\in\Om_{x}$, we have,
\begin{equation}
\label{rn}
\f_{\om}(x)=\psi_{\om}(E_{\om}(x))\, .
\end{equation}

By considering linear combinations with rational coefficients, we can
select a measurable set $F\subset \Om_{0}$ of full $\m$--measure and a
dense subset $\ga_{0}\subset\ga$ of localized operators such
that (\ref{rn}) continues to be true on $F$, for each  element of
$\ga_{0}$.

Consider now a sequence $x_{n}\in\ga_{0}$ converging to $x\in\ga$.
If $\om\in F$ we obtain
$$
\f_{\om}(x)=\lim_{n}\f_{\om}(x_{n})=\lim_{n}\psi_{\om}(E_{\om}(x_{n}))
=\psi_{\om}(E_{\om}(x))\, ,
$$
that is (\ref{rn}) holds on $F\subset\Om_{0}$, simultaneously for each 
$a\in\ga$.
\end{proof}

An immediate consequence of Theorem \ref{dis} and Proposition 
\ref{stce}, is that, according to our assumptions, a Markov state is 
automatically even.

We pass to a reconstruction theorem which parallels the analogous one 
in \cite{AF1}.

We start by choosing a subset $\G\subset I\backslash\{j_{+}\}$ 
together with a classical Markov process on $\Om$ given in 
\eqref{compact} with the Markov measure $\m_{k}$ on $\Om_{k}$ determined by the 
distributions $\pi^{j}_{\om_{j}}$ at place $j$ and the transition matrices 
$\pi^{j}_{\om_{j}\om_{j+1}}$. For each $\om$, form, according to the 
prescription (iv) above, an even one--step or 
two--step product state $\psi_{\om}$ on $\ga_{I\backslash\G}$ 
depending only on the boundaries $\om_{r_{k}}$, $\om_{l_{k+1}}$, of 
the decomposition of $\G$ into connected intervals (as before, the subscript $k$ 
describes such a decomposition). Such states are well defined, taking 
into account Theorem 1 of \cite{AM3}. Moreover, the map given as in 
\eqref{mesur} is measurable in the weak--$*$ topology.

Define $\psi\in\cs(\ga)$ as
\be
\lb{didec1}
\psi:=\int^{\oplus}_{\Om}\psi_{\om}(\ce(\,\cdot\,)(\om))\m(\di\om)\,.
\ee

Consider, for each $n\in I\backslash\{j_{+}\}$ the $\Th$--invariant 
conditional expectation 
$$
\ce_{n}:\ga_{n+1]}\mapsto\car(\ce_{n})\subset\ga_{n]}
$$ 
uniquely determined by setting for $x\in\ga_{n-1]}$, 
$x_{n}\in\ga_{\{n\}}$, $x_{n+1}\in\ga_{\{n+1\}}$,
$$
\ce_{n}(xx_{n}x_{n+1}):=x\psi(x_{n}x_{n+1})
$$
if the two--step factor $\psi(x_{n}x_{n+1})$ appears in the 
decomposition of $\psi$, or $n=l_{k}-1$ ($l_{k}$ being the left 
boundary of some interval of $\G$) and $\psi_{\om}$ depends on $\om_{l_{k}}$;
$$
\ce_{n}(xx_{n}x_{n+1}):=xx_{n}\psi(x_{n+1})
$$
if the one--step factor $\psi(x_{n+1})$ appears in the 
decomposition of $\psi$, or $n=r_{k}+1$ ($r_{k}$ being the right 
boundary of some interval of $\G$) and $\psi_{\om}$ depends on 
$\om_{r_{k}}$;
$$
\ce_{n}(xx_{n}x_{n+1}):=x\sum_{\om_{n}=1}^{2}\tr{}_{\ga_{\{n\}}}(x_{n}P^{n}_{\om_{n}})
\frac{\psi(P^{n}_{\om_{n}}x_{n+1})}{\psi(P^{n}_{\om_{n}})}P^{n}_{\om_{n}}
$$
if $n\in\G$.
\begin{thm}
\lb{ricthm}
Let $\psi\in\cs(\ga)$ in \eqref{didec1} be constructed by the 
prescriptions listed above. Then it is a Markov state w.r.t. the 
sequences $\{\ce_{n}\}_{j_{-}\leq 
n<j_{+}}$ of the above mentioned conditional expectations.
\end{thm}
\begin{proof}
A straighforward computation, taking into account the various 
possibilities, see the analogous proof of Theorem 4.1 of \cite{AF1}.
\end{proof}

\section{connection with statistical mechanics}

In this section we investigate natural connections between the Markov 
property and the KMS conditions for states on CAR algebra. This 
provides natural applications to quantum statistical mechanics, 
see \cite{AF1}--\cite{AM2}, for other analogous connections.

Suppose we have a locally faithful state on the CAR
algebra $\ga$, then a
potential $h_{\L}$ is canonically 
defined for each finite subset $\L$ of the index set $I$ by
\begin{equation}
\label{is}
\f_{\lceil\ga_{\L}}=\tr{}_{\ga_{\L}}(e^{-h_{\L}}\,\cdot\,)\, .
\end{equation}

Such a set of potentials $\{h_{\L}\}_{\L\subset I}$ satisfies normalization
conditions
\begin{equation*}
\tr{}_{\ga_{\L}}(e^{-h_{\L}})=1\, ,
\end{equation*}
together with compatibility conditions
\begin{equation*}
(\tr{}_{\gb_{\L}}\otimes\id{}_{\ga_{\L}})
(e^{-h_{\widehat{\L}}})=e^{-h_{\L}}
\end{equation*}
for finite subsets $\L\subset\widehat{\L}$, whenever 
$\ga_{\widehat{\L}}\cong\gb_{\L}\otimes\ga_{\L}$.

As the structure of Markov states is fully understood, the set of
potentials related to $\f$ by (\ref{is}) can be esplicitely written and
satisfies some nice properties.

We start by defining sequences of selfadjoint matrices 
$\{H_{j}\}_{j_{-}\leq j\leq j_{+}}$,\\
$\{\widehat{H}_{j}\}_{j_{-}\leq j\leq j_{+}}$
localized in $\ga_{\{j\}}$, and
$\{H_{j,j+1}\}_{j_{-}\leq j<j_{+}}$ localized in
$\ga_{[j,j+1]}$ respectively. Let the distribution 
$\pi^{j}_{\om_{j}}$ at place $j$, and the transition coefficients 
$\pi^{j}_{\om_{j}\om_{j+1}}$ be defined in \eqref{trp}.
Denote by $\r_{\psi}$ the density--matrix associated to a 
strictly positive functional $\psi$ on a full matrix algebra. 
If $\ve_{j}=\ga_{\{j\},+}$, define for $x\in\ga_{\{j-1\}}$ and 
$y\in\ga_{\{j+1\}}$, 
$l_{\om_{j}}(x):=\f(xP^{j}_{\om_{j}})$, 
$r_{\om_{j}}(x):=\f(P^{j}_{\om_{j}}y)$. The form $l$, $r$ are positive 
functionals on $\ga_{\{j-1\}}$, $\ga_{\{j+1\}}$ respectively.
Put
\begin{align}
\lb{hmlt}
&H_{j}=0\,\,, \widehat{H}_{j}=-\ln\r_{\f\lceil_{\ga_{\{j\}}}}\,,
\quad\car(\ve_{j})=\bc I\,;\nn\\
&H_{j}=-\ln\r_{\f\lceil_{\ga_{\{j\}}}}\,\,,\widehat{H}_{j}=0\,,
\quad\car(\ve_{j})=\ga_{\{j\}}\,;\nn\\
&H_{j}=-\sum_{\om_{j}}\big(\ln\pi^{j}_{\om_{j}}\big)P^{j}_{\om_{j}}\,\,,
\widehat{H}_{j}=0\,,
\quad\car(\ve_{j})=\ga_{\{j\},+}\,;\nn\\
&H_{j,j+1}=-\ln\r_{\f\lceil_{\ga_{\{j\}}}}\,,
\quad\car(\ve_{j})=\car(\ve_{j+1})=\bc I\,;\nn\\
&H_{j,j+1}=-\ln\r_{\f\lceil_{\ga_{\{j+1\}}}}\,,
\quad\car(\ve_{j})=\bc I\,\,,\car(\ve_{j+1})=\ga_{\{j+1\}}\,;\nn\\
&H_{j,j+1}=-\sum_{\om_{j+1}}\ln\r_{l_{\om_{j+1}}}P^{j+1}_{\om_{j+1}}\,,\nn\\
&\car(\ve_{j})=\bc I\,\,,\car(\ve_{j+1})=\ga_{\{j+1\},+}\,;\nn\\
&H_{j,j+1}=0\,,\quad\car(\ve_{j})=\ga_{\{j\}}\,\,,\car(\ve_{j+1})=\bc I\,;\\
&H_{j,j+1}=-\ln\r_{\f\lceil_{\ga_{[j,j+1]}}}\,,
\quad\car(\ve_{j})=\ga_{\{j\}}\,\,,\car(\ve_{j+1})=\ga_{\{j+1\}}\,;\nn\\
&H_{j,j+1}=-\sum_{\om_{j+1}}\big(\ln\pi^{j+1}_{\om_{j+1}}\big)P^{j+1}_{\om_{j+1}}\,,\nn\\
&\car(\ve_{j})=\ga_{\{j\}}\,\,,\car(\ve_{j+1})=\ga_{\{j+1\},+}\,;\nn\\
&H_{j,j+1}=0\,,\quad\car(\ve_{j})=\ga_{\{j\},+}\,\,,\car(\ve_{j+1})=\bc I\,;\nn\\
&H_{j,j+1}=-\sum_{\om_{j}}\ln P^{j}_{\om_{j}}\r_{r_{\om_{j}}}\,,\nn\\
&\car(\ve_{j})=\ga_{\{j\},+}\,\,,\car(\ve_{j+1})=\ga_{\{j+1\}}\,;\nn\\
&H_{j,j+1}=-\sum_{\om_{j},\om_{j+1}}\big(\ln\pi^{j}_{\om_{j}\om_{j+1}}\big)
P^{j}_{\om_{j}}P^{j+1}_{\om_{j+1}}\,,\nn\\
&\car(\ve_{j})=\ga_{\{j\},+}\,\,,\car(\ve_{j+1})=\ga_{\{j+1\},+}\,.\nn
\end{align}

Such opertors are even, and it is easy to verify that they satisfy the following
commutation relations
\begin{align}
\label{is1}
&[H_{j},H_{j,j+1}]=0,\quad
[H_{j,j+1},\widehat{H}_{j+1}]=0\, ,\nn\\
&[H_{j},\widehat{H}_{j}]=0,\quad [H_{j,j+1},H_{j+1,j+2}]=0\,.
\end{align}
\begin{thm}
Let $\f\in\cs(\ga)$ be a (locally faithful) Markov state.

Then the pointwise norm--limit
$$
\lim_{\stackrel{k\to j_{-}}{l\to
j_{+}}}e^{-ith_{[k,l]}}ae^{ith_{[k,l]}}
$$
exists and defines a one--parameter automorphisms group $t\mapsto\s_{t}$ on the
CAR algebra $\ga$ which admits $\f$ as a KMS state.
Further, $\f$ has
a normal faithful extension on all of $\pi_{\f}(\ga)''$. 
In particular, any Markov state is faithful.
\end{thm}
\begin{proof}
We start by noticing that, for each $k\leq l$,
\begin{equation}
\label{is2}
h_{[k,l]}=H_{k}+\sum_{j=k}^{l-1}H_{j,j+1}+\widehat{H}_{l}.
\end{equation}

Here, $h_{[k,l]}$ is the potential of $\f$ relative to the region 
$[k,l]$ according to \eqref{is}, and the even selfadjoint operators 
$H_{k}$, $H_{j,j+1}$, $\widehat{H}_{l}$ are given in 
\eqref{hmlt} and satisfy the commutation relations \eqref{is1}. 
Thanks to these properties, the cocycle
$e^{ith_{[k-1,l+1]}}e^{-ith_{[k,l]}}$ commutes with each element
$a\in\ga$ localized in $\ga_{[k+1,l-1]}$.
Then
$e^{-ith_{[k,l]}}ae^{ith_{[k,l]}}$ becomes
asymptotically constant
($t$ fixed) on the localized elements $a\in\ga$, that is it trivially
converges, pointwise
in norm, on the localized elements of $\ga$. Next, by a standard
$3$--$\e$ trick, it converges on all of $\ga$ and defines an
isometry $\s_{t}$.
It is straigthforward to show that $t\mapsto\s_{t}$ is actually a
group of automorphisms of $\ga$, which is also
pointwise--norm continuous in $t$, that is a
strongly continuous group of automorphisms of $\ga$. By constuction, $\f$ is
automatically a KMS state for $\s_{t}$ at inverse temperature 
$\b=-1$.\footnote{For the definition of Kubo--Martin--Schwinger 
boundary condition, as well as its connections with operator algebras 
and its meaning in quantum statistical mechanics, see \cite{BR2} and 
the references cited therein.}
The last assertions follow by \cite{BR2},
Corollary 5.3.9, taking into account that $\ga$ is a
simple $C^{*}$--algebra (\cite{BR1}, Proposition 2.6.17).
\end{proof}

\section{some illustrative examples}

In this section we describe some natural examples of Markov states on 
the CAR algebra. We consider the case $I=\bz$ for the index set $I$. 

We start by considering the case when the range of the two--step 
conditional expectations $\ve_{n}$ are always equal to $\bc I$, 
$\n\in I$, or 
$\ga_{\{n\}}$, $\n\in I$. In this situation, it is immediate to show (by Theorem 
\ref{dis} or by direct computation) that the Markov state $\f$ is the 
one--step product state extension of its restrictions to one--site  
local algebras:
$$
\f(x_{k}\cdots x_{l})=\f\lceil_{\ga_{\{k\}}}(x_{k})\cdots
\f\lceil_{\ga_{\{l\}}}(x_{l})\,.
$$

In this situation, $\f$ is translation--invariant iff 
$\f\lceil_{\ga_{\{n\}}}=\f\lceil_{\ga_{\{n+1\}}}\circ\a$, where $\a$ 
is the one--step (right) shift on the chain. The Hamiltonian, which 
does not contains interaction terms, is easily written taking into 
account that it is a one--step product state.

Consider the case when the range of the two--step 
conditional expectations $\ve_{n}$ are all equal to  
$\ga_{\{n\},+}$. Then, $\G=\bz$ and $\ce$ is the trace--preserving 
conditional expectation onto the maximal Abelian subalgebra $\gd\sim 
C(\Om)$ generated 
by $P^{n}_{1}\equiv a_{n}a^{+}_{n}$ and $P^{n}_{2}\equiv 
a^{+}_{n}a_{n}$, $n\in\bz$. Under our definition, if $x\in\ga$, $\ce(x)$ is 
represented by a continuous complex--valued function on $\Om$. Hence, 
we obtain
$$
\f(x)=\int_{\Om}\ce(x)(\om)\m(\di\om)\,.
$$

Notice that, in this situation, the Markov state under consideration 
is the diagonal lifting to all of $\ga$, of the classical Markov process on $\gd$ 
obtained by $\f\lceil_{\gd}$. 

The Markov state is translation invariant iff the underlying Markov measure 
$\m$ on ${\ds\Om\equiv\prod_{\bz}\{1,2\}}$ is translation invariant, 
that is iff the transition coefficients
$\frac{\f(P^{j}_{k}P^{j+1}_{l})}{\f(P^{j}_{k})}=:\pi_{\om_{j}\om_{j+1}}$ does not 
depend on $j\in\bz$, and the all distributions coefficients 
$\f(P^{j}_{\om_{j}})=:\pi_{\om_{j}}$ at places $j$ coincide with the unique 
stationary distribution 
for the primitive matrix 
$\pi:=[\pi_{\om_{j}\om_{j+1}}]$. Such a Markov state is the natural 
generalization of the Ising model to the CAR algebra.

The Hamiltonian for this Ising--like example is easily written taking 
into account that it is a diagonal lifting of a classical Markov 
chain, see \eqref{hmlt}. We report it for the sake of convenience. 
\begin{align*}
&H_{j}=-\sum_{\om_{j}}\big(\ln\pi^{j}_{\om_{j}}\big)P^{j}_{\om_{j}}\,,
\quad\widehat{H}_{j}=0\,,\\
&H_{j,j+1}=-\sum_{\om_{j},\om_{j+1}}\big(\ln\pi^{j}_{\om_{j},\om_{j+1}}\big)\,.
P^{j}_{\om_{j}}P^{j+1}_{\om_{j+1}}\,,
\end{align*}
\begin{thm}
\lb{mifa}
The translation invariant Markov state in the situation when 
$\car(\ve_{n})=\ga_{\{n\}}$, $n\in I$ is exponentially mixing w.r.t. 
the spatial translations. 
Moreover, it is a factor state.
\end{thm}
\begin{proof}
Let $k\leq l<m\leq n$ and $x\in\ga_{[k,l]}$, $y\in\ga_{[m,n]}$, we 
compute, taking into account that the functions on $\Om$ 
representing $\ce(x)$, $\ce(y)$, depend only on variables localized 
in $[k,l]$, $[m,n]$ respectively,
\begin{align*}
&\f(xy)=\sum_{\om_{k},\dots,\om_{n}}\pi_{\om_{k}}\pi_{\om_{k}\om_{k+1}}
\cdots\pi_{\om_{n-1}\om_{n}}\ce(x)(\om_{k},\dots,\om_{l})
\ce(y)(\om_{m},\dots,\om_{n})\\
&=\sum_{\hskip-5pt{\renewcommand\arraystretch{.7}\begin{array}c
\scriptstyle \om_{k},\dots,\om_{l}\\
\scriptstyle \om_{m},\dots,\om_{n}\end{array}}}
\pi_{\om_{k}}\pi_{\om_{k}\om_{k+1}}
\cdots\pi_{\om_{l-1}\om_{l}}(\pi^{m-l})_{\om_{l}\om_{m}}
\pi_{\om_{m}\om_{m+1}}
\cdots\pi_{\om_{l-1}\om_{l}}\\
&\times\ce(x)(\om_{k},\dots,\om_{l})
\ce(y)(\om_{m},\dots,\om_{n})\\	
&\rightarrow_{m-l\to+\infty}
\sum_{\om_{k},\dots,\om_{l}}\pi_{\om_{k}}\pi_{\om_{k}\om_{k+1}}
\cdots\pi_{\om_{l-1}\om_{l}}\ce(x)(\om_{k},\dots,\om_{l})\\
&\times\sum_{\om_{m},\dots,\om_{n}}\pi_{\om_{m}}\pi_{\om_{m}\om_{m+1}}
\cdots\pi_{\om_{n-1}\om_{n}}\ce(y)(\om_{m},\dots,\om_{n})=\f(x)\f(y)\,.
\end{align*}

Here, the exponential rate of convergence follows taking into account 
that the $r$--power $\pi^{r}$ of the primitive matrix 
$\pi$ tends exponentially to the stochastic projection onto the 
one--dimentional subspace generated by the stationary distribution 
for $\pi$, see 
e.g. \cite{S}, Section I.9. 
The fact that the Markov state $\f$ under consideration is a factor state, 
can be proved as follows. Namely, for $k=1,2,\dots$ define 
$$ 
k:= \left\{
\begin{array}{ll} 2j+1\,,\quad   j\geq0\,,\,\,k\,\text{odd}\,,\\ 
-2j\,,\quad\quad   j<0\,,\,\,k\,\text{even}\,, \end{array}\right.
$$
apply to the ordered set $k=1,2,\dots$ the construction of pag. 92 
of \cite{T3} concerning the set 
$\big\{\{e_{mn}(k)\}_{m,n=1}^{2}\,\,\big|\,\,k=1,2,\dots\big\}$,
and consider the new local structure generated by the algebras
$$
\gb_{\{j\}}:=\text{span}
\big\{e_{mn}(k(j))\,\,\big|\,\,m,n=1,2\big\}\,.
$$

Obiously,
\begin{itemize}
\item[(i)] $\ga_{\{j\},+}\subset\gb_{\{j\}}\,,\,j\in\bz$,
\item[(ii)] $\big[\gb_{\{j_{1}\}},\gb_{\{j_{2}\}}\big]=0\,,\,\, j_{1}\neq j_{2}\,,\,
j_{1},j_{2}\in\bz$,
\item[(iii)] ${\ds\overline{\bigvee_{j\in\bz}\gb_{\{j\}}}^{C^{*}}=\ga}$.
\end{itemize}

The last assertion directly follows from Theorem 2.6.10 of 
\cite{BR1}, by applying the previous considerations about the 
clustering to the new filtration $\big\{\gb_{\{j\}}\big\}_{j\in\bz}$.
\end{proof}
	
Other interesting examples are the two--block 
factors (see \cite{Z} for the analogy with the classical situation). 
These (two) examples arise when the ranges of 
two--point expectations are alternatively $\bc I$ and $\ga_{\{\,\cdot\,\}}$, 
say, $\car(\ve_{2n})=\bc I$ and 
$\car(\ve_{2n+1})=\ga_{\{2n+1\}}$. In the last situation, we get
$$
\f(x_{2k}x_{2k+1}\cdots x_{2l}x_{2l+1})=\f\lceil_{\ga_{[2k,2k+1]}}(x_{2k}x_{2k+1})
\cdots\f\lceil_{\ga_{[2l,2l+1]}}(x_{2l}x_{2l+1})\,,
$$
that is it is the two--point product state extension. It is two--step 
translation invariant iff 
$\f\lceil_{\ga_{\ga_{[2n,2n+1]}}}=\f\lceil_{\ga_{[2n+2,2n+3]}}\circ\a^{2}$, $\a$ 
being the shift on the chain.

The Hamiltonian for this two--block factor, which is a particular case 
of those described in \eqref{hmlt}, is easily written as follows. 
\begin{align*}
&H_{2j,2j+1}=-\ln\r_{\f\lceil_{\ga_{[2j,2j+1]}}}\,,\quad 
H_{2j+1,2j+2}=0\,,\\
&H_{2j}=\widehat{H}_{2j+1}=0\,,\quad
H_{2j+1}=-\ln\r_{\f\lceil_{\ga_{\{2j+1\}}}}\,,\quad
\widehat{H}_{2j}=-\ln\r_{\f\lceil_{\ga_{\{2j\}}}}\,.
\end{align*}

The Hamiltonian for the other example of two--block factor is written 
in a similar way.

To end the section, the following remark is in order. 
By applying Theorem \ref{mifa} and Proposition 3 of \cite{R}, 
it is immediate to show that all the other states considered in this 
section, as well as the family $\{\psi_{\om}\circ\ce\}_{\om\in\Om}$ appearing 
in \eqref{didec} (equivalently in \eqref{didec1}), denoted 
symbolically by $\eta$, with GNS representation $\pi_{\eta}$, 
provide examples for which the {\it algebra at 
infinity} $\gz^{\perp}_{\pi_{\eta}}$ is trivial.\footnote{See 
\cite{BR1}, Definition 2.6.4 for the definition of the algebra at 
infinity.} One could conclude that the states 
$\eta$ are factor states if he is able to prove the inclusion 
$\gz_{\pi_{\eta}}\subset\pi_{\eta}(\ga_{+})''$, see the remark after 
Proposition 4 of \cite{R}. Unfortunately, the last inclusion is false 
in general, see \cite{MV}.

\section{construction of Markov states}

In this section we are going to demonstrate concrete constructions
of Markov states. In the sequel we will assume that for the index
set $I=\bz_-$. According to Proposition \ref{twst} it is enough to
construct a functional $\f$ on $\ga$, which is a Markov state with
respect to the quasi-conditional expectation $E_{n]}$.

By $\ce_{n]}$ ( resp. $\ce_{[n}$) we denote
$\ce_{n]}:\ga\to\ga_{n]}$ (resp. $\ce_{[n}:\ga\to\ga_{[n,0]}$ ),
here $n\in\bz_-$, the canonical Umegaki conditional expectation
with respect to the trace. Note that the existence of such
expectations have been proven by Araki and Moriya in \cite{AM2}.

Let be given an even positive operator $w_0\in \ga_{\{0\},+}$ and
a sequence of even operators $\{K_{n-1,n}\}\subset\ga_{[n-1,n],+}$

\begin{defin}\label{cda} We say that the sequence $\{K_{n-1,n}\}$ 
describes
a sequence of conditional density amplitudes if it satisfies the
following conditions
\begin{itemize}
\item[(i)] $\ce_{n-1]}(K_{n-1,n}K_{n-1,n}^*)=\id \ \ \ \ n\leq
-2;$
\item[(ii)] $\ce_{-1]}(K_{-1,0}w_0K^*_{-1,0})=\id$; \\
\item[(iii)] $\ce_{[n}(K_{n-1,n}^*K_{n-1,n})=\id \ \ \ n\leq
-1.$
\end{itemize}
\end{defin}
Denote
$$
{\mathbf{K}}_{n-1}=K_{n-1,n}\cdots K_{-1,0}w_0^{1/2} \ \ \ \
{\mathbf{K}}_{n-1,k}=K_{n-1,n}\cdots K_{k-1,k}, \ \ \ n<k.
$$

For $n\in\bz_-$ put
$$
W_{[n,0]}={\mathbf{K}}_n^*{\mathbf{K}}_n.
$$

Define $E_{n]}:\ga\to\ga_{n]}$ as follows \be \lb{conex1}
E_{n]}(x)=\ce_{n]}({\mathbf{K}}_nx{\mathbf{K}}^*_n), \ \ \
x\in\ga. \ee

 From Corollary 4.8 \cite{AM2} we infer the following

\begin{lem}
\lb{conex2} For every $n\in\bz_-$ the equality holds
$$
\Th E_{n]}=E_{n]}\Th.
$$
\end{lem}

Since Umegaki conditional expectation is completely positive,
therefore we have

\begin{lem}\lb{cp} The map $E_{n]}$ defined by \eqref{conex1} is 
completely positive.
\end{lem}

Recall that a family $\{F_{[n,0]}\}$, where
$F_{[n,0]}\in\ga_{[0,n]}$, is called {\it projective} if
\be\lb{pro1}\ce_{[n}(F_{[n-1,0]})=F_{[n-1,0]} \ee is valid for all
$n\leq -1$.

\begin{lem}\lb{pro2} The family $\{W_{[n,0]}\}$ is a projective family
of density matrices.
\end{lem}

\begin{proof} Using (iii) of Def.\ref{cda} we check \eqref{pro1}:
\begin{align*}
\ce_{[n}(W_{[n-1,0]})=&\ce_{[n}(w_0^{1/2}K^*_{-1,0}K^*_{-2,-1}\cdots
K^*_{n-1,n}K_{n-1,n}\cdots K_{-1,0}w_0^{1/2})\\
=&w_0^{1/2}K^*_{-1,0}K^*_{-2,-1}\cdots
\ce_{[n}(K^*_{n-1,n}K_{n-1,n})\cdots K_{-1,0}w_0^{1/2}\\
=& w_0^{1/2}K^*_{-1,0}K^*_{-2,-1}\cdots
K^*_{n,n+1}K_{n,n+1}\cdots K_{-1,0}w_0^{1/2}\\
=&W_{[n,0]}.
\end{align*}

Finally, condition (ii) of Def.\ref{cda} implies that such
$W_{[n,0]}$ is a density matrix.
\end{proof}

Let  $\t_{[k,n]}$ be the normalized trace on $\ga_{[k,n]}$. Define
a functional on $\ga_{[n,0]}$ as follows
$$
\f_{[n,0]}(x)=\t_{[n,0]}(W_{[n,0]}x), \ \ \ x\in \ga_{[n,0]}.
$$
Then using a property of Umegaki conditional expectations we infer
that
$$
\f_{[n,0]}(x)=\t_{[n,0]}({\mathbf{K}}_nx{\mathbf{K}}^*_n)=\t_{\{n\}}
(\ce_{\{n\}}({\mathbf{K}}_nx{\mathbf{K}}^*_n)),
$$
here $\ce_{\{n\}}:\ca_{n]}\to\ca_{\{n\}}$ is a Umegaki conditional
expectation. According to Theorem 4.7 \cite{AM2} we have
$$
\ce_{n]}\upharpoonright_{\ga_{[n}}=\ce_{\{n\}}
$$
therefore
$$
\f_{[n,0]}(x)=\t_{\{n\}}(\ce_{n]}({\mathbf{K}}_nx{\mathbf{K}}^*_n))
$$

According to Lemma \ref{pro2} and (i)-(ii) of Def.\ref{cda} we
conclude that such functionals are compatible family of states. So
we can extend such states $\f_{[n,0]}$ to  $\ga$, which is denoted
by $\f$. From Lemma \ref{conex2} we conclude that $\f$ is
$\Th$-invariant.

\begin{thm}\lb{marc} The functional $\f$ is a Markov state.
\end{thm}

\begin{proof} From \eqref{conex1} and properties of the Umegaki 
conditional expectations
one can see that the maps $E_{n]}$ are quasi-conditional
expectations with respect to the triple
$(\ga_{n-1]},\ga_{n]},\ga)$, and it is easy to check that
$E_{n]}(\ga_{[n})\subset\ga_{\{n\}}$.

Since the operators $\{K_{n-1,n}\}_{n\in\bn}$ and $w_0$ are even,
therefore that states $\{\f_{[n,0]}\}_{n\in\bn}$ are even, hence
also $\f$ is so.

Denote by $\f_{[n,k]}$ the restriction of the state $\f$ on
$\ga_{[n,k]}$. Let us find a density of the this state. Using the
evenness of $K_{m,m+1}$ and (i)-(iii) of Def.\ref{cda}  we get
$$
\f_{[n,k]}(x_n\cdots x_k)=\t_{[n,0]}({\mathbf{K}}_n(x_n\cdots
x_{k}){\mathbf{K}}^*_n)=\t_{[n,k]}({\mathbf{K}}_{n,k}x_n\cdots
x_k{\mathbf{K}}_{n,k}^*).
$$

Now we are going to  check the first condition of the item (iii)
of Proposition \ref{twst}. To show one it is enough to verify the
following equality
$$
\f_{[n,h-1]}(E_{h]}(x))=\f_{[n,0]}(x), \ \ \ x\in\ga_{[n,0]}
$$
We have
\begin{eqnarray*}
\f_{[n,h-1]}(E_{h]}(x))&= & \t_{[n,h-1]]}({\mathbf{K}}_{n,h-1}
E_{h]}(x){\mathbf{K}}_{n,h-1}^*)\\
&=
&\t_{[n,h-1]}({\mathbf{K}}_{n,h-1}\ce_{h]}({\mathbf{K}}_{h}x
{\mathbf{K}}_{h}^*){\mathbf{K}}_{n,h-1}^*)\\
&=&\t_{[n,h-1]}(\ce_{h]}({\mathbf{K}}_{n,h-1}{\mathbf{K}}_{h}x
{\mathbf{K}}_{h}^*{\mathbf{K}}_{n,h-1}^*))\\
&=& \t_{[n,h-1]}(E_{n]}(x)) \\
&=&\t_{\{n\}}(E_{n]}(x))=\f_{[n,0]}(x).\\
\end{eqnarray*}

Thus $\f$ is a Markov state.
\end{proof}

 From this theorem we infer that any sequence of conditional
density amplitudes defines a Markov state. Therefore it is enough
to construct such kind of sequence to produce some concrete
examples of Markov states. Now we give certain examples of
sequences of
conditional density amplitudes. \\

{\bf Example 6.1.} Let us denote
$\tilde\e_{n}=\ce_{n-1]}\upharpoonright_{\ga_{[n-1,n]}}$ and
$\e_{n}=\ce_{[n}\upharpoonright_{\ga_{[n-1,n]}}$.

Define a sequence of operators $\{B_n\}\subset \ga_{[n-1,n]}$ as
follows
\begin{align}
\label{bn}
B_{n-1,n}(\a,\b,\g,\d)=&\a a_{n-1}^*a_{n-1}a^*_na_n+\b
a_{n-1}a^*_{n-1}a^*_na_n\\ \nn
+&\g a_{n-1}^*a_{n-1}a_na^*_n+\d
a_{n-1}a_{n-1}^*a_na^*_n,
\end{align}
where $\a,\b,\g,\d\in\br$. It is clear that from \eqref{bn} we
have that $\Th_I(B_{n-1,n})=B_{n-1,n}$ for all $I\subset \bz_-$.
This means each operator $B_{n-1,n}$ is even for all $n\leq 0$.
Therefore, the operator  $D_{n-1,n}(h)=\exp(hB_{n-1,n})$,
$h\in\br$, is positive and even for all $n\leq 0$.

 From \eqref{bn} we can easily get the equality
\begin{equation*}
(B_{n-1,n}(\a,\b,\g,\d))^k=B_{n-1,n}(\a^k,\b^k,\g^k,\d^k)
\end{equation*}
for all $k,n\geq 1$. Using this we infer
\begin{equation*}
D_{n-1,n}(h)=B_{n-1,n}(e^{h\a},e^{h\b},e^{h\g},\e^{h\d}).
\end{equation*}

Since $\tilde\e_{n}$ and $\e_{n}$ are Umegaki conditional
expectations, we have

\begin{equation*}
\tilde\e_{n}(D_{n-1,n}(h))=\frac{1}{2}\bigg((e^{h\a}+e^{h\b})
a_{n-1}^*a_{n-1}+(e^{h\g}+e^{h\d})a_{n-1}a_{n-1}^*\bigg).
\end{equation*}
\begin{equation*}
\e_{n}(D_{n-1,n}(h))=\frac{1}{2}\bigg((e^{h\a}+e^{h\b})
a_n^*a_n+(e^{h\g}+e^{h\d})a_na_n^*\bigg);
\end{equation*}

Impose that
\begin{equation*}
e^{h\a}+e^{h\b}=e^{h\g}+e^{h\d}.
\end{equation*}

Therefore, denote
\begin{equation*}
\kappa=\frac{e^{h\a}+e^{h\b}}{2}.
\end{equation*}

Whence we have
\begin{equation}\label{bn2}
\tilde\e_{n}(D_{n-1,n}(h))=\kappa\id \ \ \ \ \
\e_{n}(D_{n-1,n}(h))=\kappa \id.
\end{equation}

Put $w_0=\id$ and
$$
K_{n-1,n}=\frac{1}{\sqrt{\kappa}}D_{n-1,n}(h/2).
$$

 From \eqref{bn2} we can prove the following: for every $n\in\bz_-$
we have
\begin{equation*}
\tilde \e_{n}(K_{n-1,n}K_{n-1,n}^*)=\id.
\end{equation*}

Indeed
\begin{eqnarray}\label{kk}
\tilde\e_{n}(K_{n-1,n}K_{n-1,n}^*)&=&
\tilde\e_n(\frac{1}{\sqrt{\kappa}}D_{n-1,n}(h/2)D_{n-1,n}(h/2)
\frac{1}{\sqrt{\kappa}})=\nonumber
\\
&=&\frac{1}{\kappa}\tilde \e_n(D_{n-1,n}(h))=\id.
\end{eqnarray}

Using the same argument one can show
\begin{equation*}
\e_{n}(K_{n-1,n}^*K_{n-1,n})=\id \ \ \ \ \forall n\leq -1.
\end{equation*}

So  according to Theorem \ref{marc}  we can construct a Markov
state. Note that the constructed QMS can be interpreted as
'Fermion' Ising model, and it coincides with the second
illustrative example of  section 5. \\

{\bf Example 6.2.} Define a sequence of operators
$\{U_n\}\subset\ga_{[n-1,n]}$ as follows
$$
U_n=a_{n-1}^*a_n+a_{n}^*a_{n-1}.
$$
Put $V_{n-1,n}=\exp(hU_n/2)$, where $h\in\br$. It is clear that
each  $V_{n-1,n}$ is positive and even, since $U_n$ is even for
all $n\in\bz_-$.

Now according to Theorem 4.7 \cite{AM2} $\tilde\e_{n}$ has a form
$$
\tilde\e_{n}(a)=\ce_n^{(2)}\ce^{(1)}_n(a), \ \ \
a\in\ga_{[n-1,n]},
$$
where $\ce_n^{(2)}$ and $\ce_n^{(1)}$ are defined in \cite{AM2}.

Let us compute the powers of $U_n$. We have
$$
U_n^2=a^*_{n-1}a_{n-1}a_{n}a^*_{n}+a_{n-1}a^*_{n-1}a^*_{n}a_{n}
=:p_{n-1,n}+q_{n-1,n}
$$
where we have denoted
$$
p_{n-1,n}=a^*_{n-1}a_{n-1}a_{n}a^*_{n}, \ \ \ \
q_{n-1,n}=a_{n-1}a^*_{n-1}a^*_{n}a_{n}.
$$
It is easy to see that they are projections such that
$p_{n-1,n}\cdot q_{n-1,n}=0$. This implies that
$$
U_n^{2k}=U_n^2, \ \ \ k\geq 1
$$
and therefore
$$
U_n^{2k+1}=U_n, \ \  k\geq 1
$$
Then for $h\in\br$ we have
\begin{eqnarray*}
\exp(hU_n)=\sum_{k\geq 0}\frac{h^k}{k!}U^k_n &= & \sum_{k\geq
0}\frac{h^{2k}}{(2k)!}U^{2k}_n+\sum_{k\geq
0}\frac{h^{2k+1}}{(2k+1)!}U^{2k+1}_n\\
&=& \id+\sum_{k\geq 1}\frac{h^{2k}}{(2k)!}U^{2}_n+\sum_{k\geq
0}\frac{h^{2k+1}}{(2k+1)!}U_k\\
&=& \id+(\sin h) U_n+(\cos h-1)U_n^2\\
\end{eqnarray*}

This implies that
$$
\ce^{(1)}_n(\exp(hU_n))=(\exp(hU_n)+\Th(\exp(hU_n)))/2=\id+ (\cos
h-1)U_n^2,
$$
from this we get
\begin{eqnarray*}
\tilde\e_{n}(V_{n-1,n}^2)=\tilde\e_{n}(\exp(hU_n))&=&\id+(\cos 
h-1)\ce^{(2)}_n(U_n^2)\\
&=&\id+\frac{\cos h-1}{2}(a_{n-1}^*a_{n-1}+a_{n-1}a_{n-1}^*)\\
&=&\frac{\id+\cos h}{2}\id
\end{eqnarray*}

Put $w_0=\id$ and
$$
K_{n-1,n}=\frac{1}{\sqrt{\a}}V_{n-1,n},
$$
where $\a=\frac{1+\cos h}{2}$.

  Using the argument of \eqref{kk}
one can prove the following equalities:
\begin{eqnarray*}
&&\tilde\e_{n}(K_{n-1,n}K_{n-1,n}^*)=\id \ \ \ \ \forall n\leq
0\\
&& \e_{n}(K_{n-1,n}^*K_{n-1,n})=\id \ \ \ \ \forall n\leq -1.
\end{eqnarray*}

So  according to Theorem \ref{marc}  we can construct a Markov
state.


\begin{thebibliography}{99}

\bibitem{Z} Aaronson J., Gilat D., Keane M., de Valk V.  
{\it An algebraic construction of a class of one--dependent processes}, 
Ann. Probab. {\bf 17} (1989), 128--143.

\bibitem{A} Accardi L, {\it On noncommutative Markov property},
Funct. Anal. Appl. {\bf 8} (1975), 1--8.

\bibitem{AC} Accardi L., Cecchini C.
{\it Conditional expectations in von Neumann algebras and a theorem
of Takesaki}, J. Funct. Anal. {\bf 45} (1982), 245--273.

\bibitem{AF1} Accardi L., Fidaleo F. 
{\it Non homogeneous quantum Markov states and quantum Markov fields}, 
J. Funct. Anal. {\bf 200} (2003), 324--347.

\bibitem{AF2} Accardi L., Fidaleo F. 
{\it Quantum Markov fields}, 
Infin. Dimens. Anal. Quantum Probab. 
Relat. Top. {\bf 6} (2003), 123--138.

\bibitem{AF3} Accardi L., Fidaleo F. 
{\it Recent developments on the quantum Markov property}, 
in ``Quantum Probability and Related Topics'' Vol. {\bf XV},
ed. W. Freudenberg, 1--19 World Scientific, Singapore 2003.

\bibitem{AF4} Accardi L., Fidaleo F. 
{\it Entangled Markov chains}, Ann. Mat. Pura Appl., to appear.

\bibitem{AFr} Accardi L., Frigerio A.
{\it Markovian cocycles}, Proc. R. Ir. Acad. {\bf 83} (1983), 251--263.

\bibitem{ALi} Accardi L., Liebscher V.
{\it Markovian KMS states for one dimensional spin chains},
Infin. Dimens. Anal. Quantum Probab. Relat. Top. {\bf 2} (1999), 645--661.

\bibitem{ALV} Accardi L., Lu Y. G., Volovich I. {\it Quantum theory 
and its stochastic limit}, Springer, Berlin--Heidelberg--New
York, 2002.

\bibitem{AM1} Araki H, Moriya H.
{\it Local thermodynamical stability of Fermion lattice systems},
Lett. Math. Phys. {\bf 62} (2002), 33--45.

\bibitem{AM2} Araki H, Moriya H.
{\it Equilibrium statistical mechanics of Fermion lattice systems},
Rev. Math. Phys. {\bf 15} (2003), 93--198.

\bibitem{AM3} Araki H, Moriya H.
{\it Joint extension of states of subsystems},
Commun. Math. Phys. {\bf 237} (2003), 105--122.

\bibitem{AM4} Araki H, Moriya H.
{\it Conditional expectations relative to a product state and the 
corresponding standard potentials},
Commun. Math. Phys. {\bf 246} (2004), 113--132.

\bibitem{BR1} Bratteli O., Robinson D. W.
{\it Operator algebras and quantum statistical mechanics I},
Springer, Berlin--Heidelberg--New York, 1981.

\bibitem{BR2} Bratteli O., Robinson D. W.
{\it Operator algebras and quantum statistical mechanics II},
Springer, Berlin--Heidelberg--New York, 1981.

\bibitem{F} Fidaleo F. 
{\it Infinite dimensional entangled Markov chains}, 
Random Op. Stoch. Eq., to appear.

\bibitem{FI} Fidaleo F., Isola T. {\it Minimal conditional expectations 
for inclusions with atomic centres}, Internat. J. Math. {\bf 7} (1996), 307--327.

\bibitem{FM} Fidaleo F., Mukhamedov F.
{\it Diagonalizability of non homogeneous quantum Markov states and 
associated von Neumann algebras}, preprint 2004.

\bibitem{H} Havet J.--F. {\it Esp\'erance conditionelle minimale}, 
J. Operator Theory {\bf 24} (2000), 33--55.

\bibitem{MV} Manuceau J., Verbeure A.
{\it Non--factor quasi--free states of the CAR algebra},
Commun. Math. Phys. {\bf 18} (1970), 319--329.

\bibitem{R} Robinson D. W.
{\it A characterization of clustering states},
Commun. Math. Phys. {\bf 41} (1975), 79--88.

\bibitem{S} Schaefer H. H. {\it Banach Lattices and positive operators}, 
Springer--Verlag, Berlin Heidelberg New York 1974.

\bibitem{T1} Takesaki M.
{\it Theory of operator algebras I},
Springer, Berlin-Heidelberg-New York 1979.

\bibitem{T3} Takesaki M.
{\it Theory of operator algebras III},
Springer, Berlin-Heidelberg-New York 2003.

\end{thebibliography}
\end{document}